\documentclass[pre,superscriptaddress,twocolumn,a4paper,showpacs,nofootinbib]{revtex4-1}
\usepackage{graphics}
\usepackage{amssymb}
\usepackage{amsmath}
\usepackage{wasysym}
\usepackage{latexsym}
\usepackage{graphicx}
\usepackage{epsfig}
\usepackage{subfigure}
\usepackage{float}
\usepackage{color}
\usepackage{lineno}
\usepackage{verbatim}
\usepackage{enumitem}
\usepackage[T1]{fontenc}

\begin{document}

\title{$N$-player game formulation of the  majority-vote model of opinion dynamics}

\author{Jo\~ao P. M. Soares}
\affiliation{Instituto de F\'{\i}sica de S\~ao Carlos,
  Universidade de S\~ao Paulo,
  Caixa Postal 369, 13560-970 S\~ao Carlos, S\~ao Paulo, Brazil}

\author{Jos\'e F.  Fontanari}
\affiliation{Instituto de F\'{\i}sica de S\~ao Carlos,
  Universidade de S\~ao Paulo,
  Caixa Postal 369, 13560-970 S\~ao Carlos, S\~ao Paulo, Brazil}


\begin{abstract}
From a self-centered perspective, it can be assumed that people only hold opinions that can benefit them. If opinions have no intrinsic value, and acquire their value when held by the majority of individuals in a discussion group, then we have a situation that can be modeled as an $N$-player game.
Here we explore the dynamics of (binary) opinion formation  using a game-theoretic framework to  study  an $N$-player game version of Galam's local majority-vote model. The opinion dynamics is modeled by a stochastic imitation dynamics in which the individuals copy the opinion of more successful peers. In the infinite population limit, this dynamics is  described by the  classical replicator equation of evolutionary game theory. The equilibrium solution shows a threshold separating the initial frequencies that lead to the fixation of one opinion or the other.   A comparison with Galam's deterministic model reveals contrasting results, especially in the presence of inflexible individuals, who never change their opinions. In particular, the $N$-player game predicts a polarized equilibrium consisting only of extremists.  Using finite-size scaling analysis, we evaluate the critical exponents  that  determine the population size dependence of the opinion's fixation probability and mean fixation times near the threshold.   The results underscore the usefulness of combining evolutionary game theory with opinion dynamics and  the importance of statistical physics tools to summarize  the results of  Monte Carlo simulations.
\end{abstract}

%\begin{keyword}
%Synchronization; Delay differential equations;  Fixed-point analysis;  Stochastic processes.
%\end{keyword}

\maketitle

%
%-----------------------------------------------------
\section{Introduction} \label{sec:intro}
%-----------------------------------------------------
 %
 
Conformity is probably the main psychological mechanism behind the social pressure that causes people to change their minds \cite{Asch_1955}. In fact, classical mathematical  models of opinion dynamics, such as the majority vote model \cite{Liggett_1985} (see \cite{Castellano_2009,Galam_2012,Sznajd_2019} for reviews), incorporate conformity through a frequency-dependent bias mechanism by which individuals are disproportionately likely to adopt the more common opinion among their peers \cite{Boyd_1985}. From a self-centered perspective, where holding one opinion or another has a benefit or cost to individuals, which  is ultimately the reason they hold a particular opinion, opinion dynamics may be best described within a game theoretic framework, as is done for behavioral strategies that lead to human cooperation \cite{Perc_2017,Wang_2023}, for instance.

Accordingly, we propose here an $N$-player game version of Galam's local majority-vote model \cite{Galam_2012}.   In this well-established deterministic model of (binary) opinion dynamics, individuals in an infinite population hold opinions A or B and change their opinions if they are in the minority in a randomly assembled discussion group of size $N$.    In our game version of this model, each individual's payoff is determined by randomly choosing $N-1$ other individuals to form a play group of size $N$.  If the individual's opinion happens to be the majority in her group, she is assigned payoff $w$, otherwise she is assigned payoff $l < w$. Ties require an additional rule to decide whether the   payoff is $w$ or $l$. If an individual's payoff is less than the payoff of a randomly chosen peer, she will change her opinion with a probability proportional to the payoff difference. This stochastic imitation dynamics  is  described by the replicator equation \cite{Hofbauer_1998,Nowak_2006} in the limit of infinite population size  \cite{Traulsen_2005,Sandholm_2010}.

The game theoretic formulation explicitly accounts  for  the discomfort of being in the minority by  assigning  a small payoff  to individuals in this situation, as well as for  the tendency of individuals to copy others who appear to be better off. Of course, these ingredients are implicit in Galam's  local majority model.

We find that, apart from a trivial rescaling of the time variable in the replicator equation, the results do not depend on the choice of the payoff values, provided that $w >l$.  Although the equilibrium solutions for the infinite population limit are the same for Galam's model and the $N$-player game version (i.e., the all-A and all-B fixed points), the unstable fixed point that determines which opinion will eventually become fixed in the population is different when there is a bias toward opinion A in the case of ties. More precisely, 
this unstable fixed point yields  the lower value of the initial frequency of opinion A that guarantees its fixation in the equilibrium regime. Thus, as the initial frequency of opinion A increases from $0$ to $1$, there is a threshold phenomenon that separates the regime where opinion $B$ is fixed from the regime where opinion $A$ is fixed.
This finding emphasizes that the models produce different results even in the simpler scenario where the population consists only of floaters. Floaters are individuals who may change their opinions for a prospective better payoff.  When inflexible individuals or extremists, i.e. individuals who never change their minds, are allowed, the models predict very different outcomes. In particular, the $N$ player game predicts a polarized equilibrium consisting only of extremists, while in Galam's  model the frequency of extremists  is constant.

We also carry out an extensive finite-size scaling analysis of the stochastic imitation dynamics near the deterministic threshold.  This analysis yields the scaling functions that describe the dependence of the fixation probability of opinion A, as well as the mean fixation time of any of the opinions, on the population and play group sizes.  We find that there are non-trivial critical exponents that describe how these quantities are affected by an increase in the size of the population in the threshold region. The powerful summary of simulation results provided by the finite-size scaling analysis demonstrates the usefulness of statistical physics in addressing interdisciplinary problems.

 The rest of this paper is organized as follows. In Section \ref{sec:Galam} we present an overview of Galam's local majority-vote model \cite{Galam_2012} and write down the recursion equation for the frequency of opinion A in the infinite population.  In Section \ref{sec:model} we present the stochastic $N$-player game version of Galam's model, which implements the imitation dynamics in a finite population. In Section \ref{sec:det}, we write the replicator equation that describes the imitation dynamics in the infinite population limit and calculate the frequency threshold that determines which opinion will eventually become fixed in the population.  In  Section \ref{sec:FP} we present the results of Monte Carlo simulations of the imitation dynamics and use finite-size scaling to characterize the equilibrium regime near the threshold. To illustrate the differences  between the $N$-player game of opinion formation and Galam's model, we study the effect of of inflexible individuals in the infinite population in Section \ref{sec:Inf}.  Finally, in Section \ref{sec:conc} we summarize our main results and present some concluding remarks.
 
%-----------------------------------------------------
 \section{Galam's model of opinion dynamics} \label{sec:Galam}
 %-----------------------------------------------------
 %
 
This brief overview focuses on a local majority model of opinion dynamics, first introduced in the study of bottom-up hierarchical voting models \cite{Galam_2000}, and then applied to a scenario where individuals are reshuffled along a series of local updates \cite{Galam_2002}. This framework leads to the formulation of general sequential probabilistic models that encompass a wide variety of models used to describe opinion dynamics \cite{Galam_2005a}. 

Here we follow, and slightly reinterpret for clarity, Ref.\ \cite{Galam_2012}'s account of the basic discrete-time local majority model.  Let $a_t$ denote the fraction of individuals holding opinion A at cycle $t$ in an infinite population. Thus, $1-a_t$ gives the fraction of individuals holding opinion B.  In each cycle, infinitely many discussion groups of size $N$ are formed from  individuals randomly selected from the infinite population, and the majority rule is applied to each of them. As a result, in each discussion  group all individuals with minority opinions change their opinions to conform to the majority.  In case of a tie, which can happen when the size $N$ of the discussion group is even, the so-called inertia principle is applied to choose one of the opinions A or B with probabilities $\kappa$ and $1-\kappa$, respectively. Thus, $\kappa > 1/2$ represents an intrinsic bias of the population towards opinion $A$, while $\kappa < 1/2$ represents a bias towards opinion B instead. The unbiased case is $\kappa= 1/2$. In this paper, we consider only discussion groups of even size $N$.  Since all discussion groups are the same size, the fraction of individuals holding opinion A at cycle $t+1$ must equal the fraction of discussion groups at cycle $t$ for which opinion A is the majority or wins in a tie.  Thus \cite{Galam_2012}
\begin{eqnarray}
a_{t+1} &  = &  \sum_{i=N/2+1}^N \binom{N}{i} a_t ^i (1-a_t)^{N-i}  \nonumber \\
&   & + \kappa  \binom{N}{N/2} a_t ^{N/2} (1-a_t)^{N/2} .
\end{eqnarray}
For $N >2$, this recursion equation has two stable fixed points, $a=0$ and $a=1$, and one unstable fixed point $a^*$, which delimits the basin of attraction of the stable fixed points. For the unbiased case $\kappa =1/2$, we have $a^*=1/2$ for all $N>2$, but $a^*$ depends on $N$ for other choices of the bias $\kappa$  
\cite{Galam_2005b}. For $N=2$ we have that $a=0$ is stable and $a=1$ is unstable if $\kappa < 1/2$, while $a=0$ is unstable and $a=1$ is stable if $\kappa > 1/2$. For $\kappa=1/2$ the recursion is frozen at the initial condition $a_0$ \cite{Galam_2012}.

A feature that distinguishes Galam's model of opinion dynamics from other majority voting models is the possibility for more than one individual to switch opinions in a discussion group, whereas in the more traditional approaches \cite{Liggett_1985,Oliveira_1992,Peres_2010,Tome_2023}, as well as in our game theoretic formulation, only one individual - the focal individual - can change opinion.

%
%-----------------------------------------------------
\section{The stochastic $N$-player game of opinion formation}\label{sec:model} 
%-----------------------------------------------------
%

Consider a well-mixed finite population of size $M$ consisting of individuals of types A and B, i.e., individuals holding mutually exclusive opinions A and B. The population is well-mixed in the sense that each individual can interact with every other individual in the population.
At each time step $\delta t$
a focal individual $i_1$  is randomly selected. The play group of $i_1$ is formed by randomly selecting $N -1$ other individuals without replacement from the remaining $M -1$ individuals in the population. The payoff  $f_{i_1}$ of focal  individual $i_1$ is $w$ (from winning) if her opinion is the majority and $l$ (from losing) if her opinion is the minority, where $w >l$.  The actual values of $w$ and $l$ are unimportant, as we will see later. In case of a tie,  the payoff of individual $i_1$ depends on her opinion. If $i_1$ is of type A, then her payoff is $w$ with probability $\kappa$ and $l$ with probability $1-\kappa$. If $i_1$ is of type B, then her payoff is $w$ with probability $1-\kappa$ and $l$ with probability $\kappa$.  As  in Galam's model, $\kappa$\ is an intrinsic bias of the population towards opinion $A$ \cite{Galam_2005b}.   Most of our analysis will consider the unbiased $\kappa=1/2$ scenario for even $N$. In the case of pairwise interactions, i.e., for  $N=2$, this game reduces to the coordination game \cite{Cooper_1998}.

Next, a model  individual $i_2 \neq i_1$  is randomly selected along with her play group, which may include focal individual $i_1$ or  other members of $i_1$'s  play group.
The payoff $f_{i_2}$ of the model individual $i_2$ is obtained using the same procedure described above.  If  $f_{i_2} > f_{i_1}$, then the focal individual $i_1$  switches to the opinion of the model individual $i_2$   with probability
\begin{equation}\label{prob0}
 \frac{f_{i_2} - f_{i_1}}{\Delta f_{\max}}, 
\end{equation}
where $\Delta f_{\max}$ is the maximum possible payoff difference that guarantees that this ratio is less than or equal to 1.  If  $f_{i_2} \leq  f_{i_1}$,  then the focal individual  $i_1$ keeps her opinion. In our game, we have $\Delta f_{\max} = w-l$, and the focal individual $i_1$  is certain to change her opinion whenever her payoff is less than the payoff of the model individual $i_2$. At this point we can already see why the actual values of $w$ and $l$ are not important in this game: all  choices of $w$ and $l$ such that $w > l$ are equivalent, since the switching probability (\ref{prob0}) is either $0$ or $1$. Thus,  in the simulations we set $w=1$ and $l=0$ without loss of generality. This completes the time step $\delta t$ of the stochastic dynamics, and so the time is updated to $t = t + \delta t$.  As usual in such an asynchronous update scheme, we choose the time increment to be $\delta t = 1/M$, so that during the increment from $t$ to $t+1$ exactly $M$, though not necessarily different, individuals are selected as focal individuals.  
%In the next section, we will give an alternative justification for this choice of time step $\delta t$. 
Our implementation of the stochastic $N$-player game is standard in the evolutionary game literature (see, e.g.,  \cite{Zheng_2007}). We refer to  \cite{Fontanari_2024} for an elementary derivation of the correspondence between the imitation dynamics and the replicator equation for the switching probability given in Eq. (\ref{prob0}),  where only more successful individuals can be imitated.

%
%-----------------------------------------------------
\section{The deterministic $N$-player game of opinion formation}\label{sec:det} 
%-----------------------------------------------------
%
Here we consider an infinitely large  population (i.e., $M \to \infty$) consisting of a fraction $x$ of individuals of type A and, consequently, a fraction $1-x$ of individuals of type B. We choose to use a different notation from Section \ref{sec:Galam} to stress that $x=x(t)$ is a function of a  real  variable, whereas $a_t$ is a  function of a discrete variable. To simplify the notation, we have used the same symbol $t$  for these two variables.  We will consider only even play group sizes $N$. The expected payoff  of an individual with opinion A is 
\begin{eqnarray}
\pi^A  (x) & = & w \sum_{i=N/2}^{N-1}  \binom{N-1}{i} x^{i} (1-x)^{N-1-i} \nonumber \\
&  &  + 
l  \sum_{i=0}^{N/2-2}  \binom{N-1}{i}  x^{i} (1-x)^{N-1-i} \nonumber \\
&  & + \left [  w \kappa + l (1-\kappa) \right ] \binom{N-1}{N/2-1} x^{N/2-1} (1-x)^{N/2}  , \nonumber  \\
\end{eqnarray}
where the first term on the right-hand side is the payoff if A is the majority, the second if A is the minority, and the third if there is a tie.  Note that since the focal  individual is of type A,  a tie occurs if there are $i = N/2 -1$ other individuals of type $A$  in the play group.  By  rearranging the terms, we obtain
\begin{eqnarray}\label{piA}
\pi^A  (x) &  =  &  w - 
(w-l)  \sum_{i=0}^{N/2-2} \binom{N-1}{i} x^{i} (1-x)^{N-1-i}  \nonumber \\
&  & -(1-\kappa) (w-l) \binom{N-1}{N/2-1} x^{N/2-1} (1-x)^{N/2} ,  \nonumber \\
\end{eqnarray}
which is maximum when $\kappa=1$, as expected. 
A similar analysis yields the expected payoff of an individual with opinion B,
\begin{eqnarray}\label{piB}
\pi^B  (x)  &  =  &  w - 
(w-l)  \sum_{j=0}^{N/2-2} \binom{N-1}{j} x^{N-1-j} (1-x)^{j} \nonumber \\
&  & - \kappa  (w-l) \binom{N-1}{N/2} x^{N/2} (1-x)^{N/2-1} ,
\end{eqnarray}
which is maximum when $\kappa=0$, as expected. 

A major advance in the theoretical understanding of public goods games, which mostly involve social learning \cite{Hardin_1968,Ostrom_1990}, is the realization that the replicator equation formalism used to study biological evolution in continuous time \cite{Hofbauer_1998,Nowak_2006} also describes the social dynamic scenario where individuals imitate the behavior of their more successful peers \cite{Traulsen_2005} (see also \cite{Sandholm_2010}).  This social dynamic  is of great importance, as it has been claimed that imitation is the fabric of human society \cite{Blackmore_2000} and the  connector of collective brains \cite{Fontanari_2014}. Of course, imitation is also at the heart of models of opinion dynamics  because, contra Mark Twain\footnote{``Whenever you find yourself on the side of the majority, it is time to pause and reflect'' \cite{Twain_2014}.},  it suggests a reason why people would change their opinions when they are on the minority side of a group: to reduce  psychological discomfort by conforming to social pressure \cite{Asch_1955}.

%-----------------------------------------------------
\begin{figure}[t] 
\centering
 \includegraphics[width=1\columnwidth]{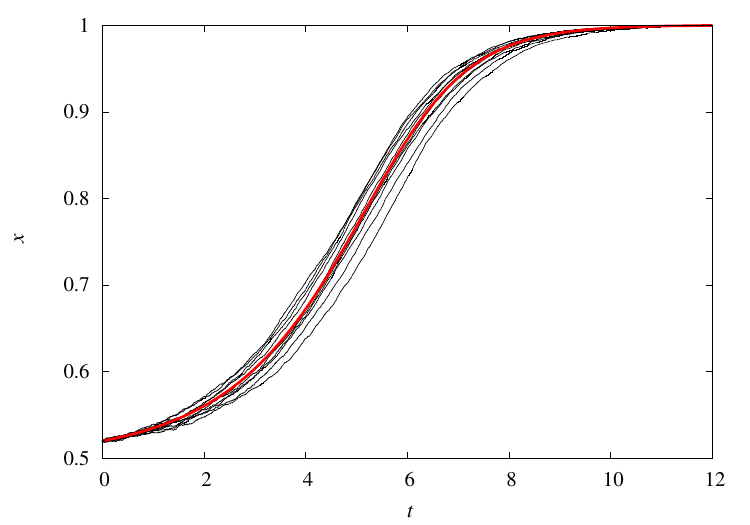}  
\caption{Fraction of  individuals with opinion A  as a function of time for play groups of size $N=4$ and initial condition $x(0) = x_0 = 0.52$. 
The jagged thin curves are single runs of the stochastic  simulation algorithm  for $M=10^4$ and the smooth thick curve is the numerical solution of the replicator equation.
 The bias parameter  is $\kappa=1/2$.
 }  
\label{fig:1}  
\end{figure}
%-----------------------------------------------------

For the $N$-player game of opinion formation, the replicator equation governing the evolution of the frequency  of type A individuals in the infinite population is
\begin{eqnarray}\label{rep1}
\frac{dx}{dt}  & = & x   \left [ \pi^A (x) - \bar{\pi}  \right ] \nonumber \\
 & = & x \left ( 1-x \right )  \left [ \pi^A (x) - \pi^B (x) \right ] ,
\end{eqnarray}
where $\bar{\pi} = x \pi^A + (1-x) \pi^B$ is the average payoff of the population and the time $t$ is measured in units of $w-l$. We refer the reader to Ref. \cite{Hofbauer_1998} for a thorough introduction to the replicator equation and to Ref. \cite{Traulsen_2005} for the connection between the stochastic imitation dynamics introduced in Section \ref{sec:model} and the replicator equation.  Figure \ref{fig:1} shows  the numerical solution of  this equation  together with  10 independent   runs of the stochastic  simulation algorithm  using a population size of $M=10^4$. The finite-size effects in the dynamics are quite significant and can be appreciated by the difference between the deterministic and the stochastic trajectories. Of course, averaging over the stochastic trajectories gives much better agreement with the deterministic prediction, but misses the finite-size effects that are averaged out.  For $M=10^5$, however,  we find a perfect agreement between the  stochastic trajectories and the solution  of the replicator equation (data not shown).

%-----------------------------------------------------
\begin{figure}[th] 
\centering
 \includegraphics[width=1\columnwidth]{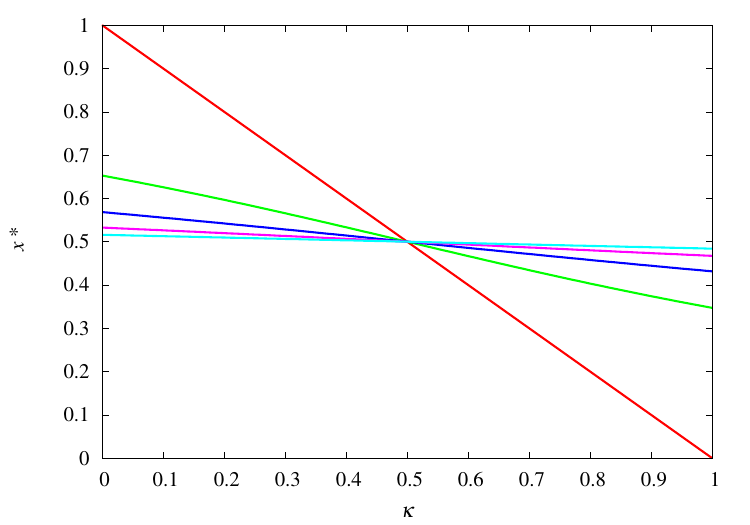}  
\caption{Unstable fixed point of the replicator equation (\ref{rep1}), which delimits the basins of attraction of the all-B ($x=0$) and all-A ($x=1$) stable fixed points, as a function of the bias towards opinion $A$ for  (from bottom to top at $\kappa=1$) $N=2,4,8,16$ and $32$. The dynamics is driven to the all-A fixed point if  $x_0 > x^*$  and to the all-B fixed point if $x_0 < x^*$.
 }  
\label{fig:2}  
\end{figure}
%-----------------------------------------------------

The equilibrium  solutions  of the replicator equation  (\ref{rep1}) are obtained by setting $dx/dt=0$. They are $x=0$ (opinion B is unanimous), $x=1$ (opinion A is unanimous) and the solution of the algebraic equation $\pi^A(x^*) =  \pi^B(x^*) $ (i.e., opinions A and B coexist) \cite{Hofbauer_1998,Nowak_2006}. The standard  linear stability analysis (see, e.g., \cite{Strogatz_2014})  shows that the fixed point $x=0$ is locally stable if  $\pi^A(0)  <  \pi^B(0) $ and that the fixed point $x=1$ is locally stable if  $\pi^A(1)  >  \pi^B(1) $. Both conditions are satisfied if $w > l$, so there is bistability.   The coexistence fixed point $x^*$  is always unstable and determines the limit of the basins of attraction of the stable fixed points: if $x_0 < x^*$,  the dynamics  leads  to $x=0$ and if $x_0 > x^*$, it leads to $x=1$. Here $x_0 \equiv x (0)$. In the unbiased case $\kappa=1/2$, we have $x^*=1/2$   regardless of the play group size $N$. For the general case, we have to solve  the $N-1$-th-order polynomial  equation  $\pi^A(x^*) =  \pi^B(x^*) $  in order to get the unstable fixed point.  Although $x^*$ is independent of $w$ and $l$, it depends on the size of the play group $N$, as shown in Fig. \ref{fig:2}.  For $N=2$ we find $x^* =1-\kappa$, which shows that the  $N$-player game of opinion formation differs from Galam's  basic  local majority model reviewed in Section \ref{sec:Galam}.
 For large $N$, where a tie is less likely, $x^*$ is not very sensitive to the value of the bias, as expected.

%
%-----------------------------------------------------
\section{Finite population simulations}\label{sec:FP} 
%-----------------------------------------------------
%
As in the deterministic limit, the ultimate outcome of the stochastic imitation dynamics for finite $M$ is the fixation of one of the opinions. 
The probability of fixation of opinion A, denoted by $\rho$, is approximated by the ratio of the number of independent runs leading to fixation of type A to the total number of runs, which we set to $10^5$ in this study.  Also of interest is the mean time $T_f$ to fixate  one  of the opinions. The leading independent variable  of this study  is  the proportion  $x_0 \in [0,1]$ of individuals of type A at the beginning of the game.  These are the quantities considered in formally similar studies of population genetics (see, e.g., \cite{Serva_2005}), where $\kappa -1/2$ can be thought of as the selective advantage (or disadvantage, if negative)  of type A individuals. The unbiased  case  $\kappa=1/2$  then corresponds to the neutral evolution scenario.

%-----------------------------------------------------
\begin{figure}[t] 
\centering
 \includegraphics[width=1\columnwidth]{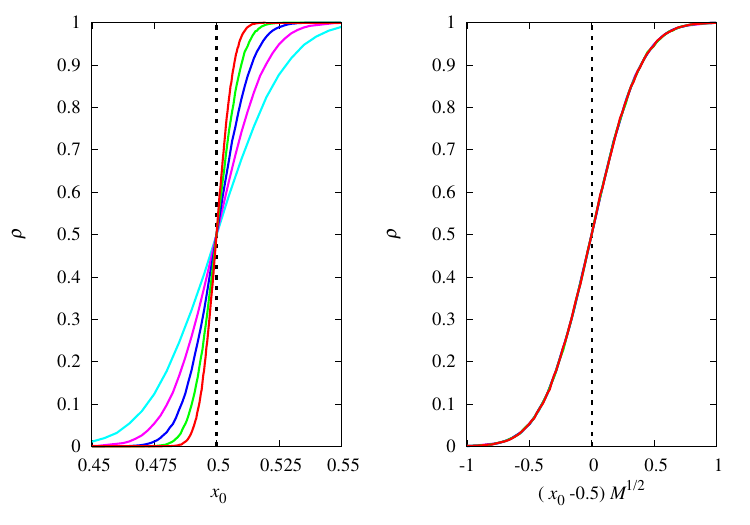}  
\caption{Probability of fixation of opinion A for play groups of size $N=4$ as a function of the initial proportion of individuals of type A (left panel) for populations of sizes (from left to right at $\rho=0.2$) $M=200, 400, 800, 1600$, and $3200$. The right panel shows $\rho$ as a function of the scaled variable $u = (x_0 - 1/2) M^{1/2}$. The vertical dashed lines indicate the threshold in the deterministic limit $M\to \infty$. The bias is $\kappa=1/2$.
 }  
\label{fig:3}  
\end{figure}
%-----------------------------------------------------

Figure \ref{fig:3} shows the probability of fixation $\rho$ for  play groups of size $N=4$ and different population sizes $M$. Recall that in the deterministic limit $M \to \infty$ we have $\rho =1$ for $x_0 > x_0^c$ and $\rho =0$ for $x_0 < x_0^c$, where $x_0^c = 1/2$ for the unbiased case $\kappa=1/2$. For finite but large $M$, we can quantify the sharpness of the threshold, i.e. the range of $x_0$ around $x_0^c$ where the threshold properties persist, using the scaling assumption \cite{Binder_1985,Privman_1990}
\begin{equation}\label{scal}
\rho =  h_N \left [ (x_0 - x_0^c)M^{1/\nu_\rho} \right ], 
\end{equation}
where $\nu_\rho > 0$ is a critical exponent and $h_N(u)$ with  $u=  (x_0 - x_0^c)M^{1/\nu_\rho}$  is the scaling function.  Note that according to the scaling assumption  $h_N(0)$  is invariant to changes in $M$, hence the procedure to determine the  threshold  $x_0^c$ as the intersection of the curves  $\rho$ {\it vs.} $x_0$ for different values of $M$.  This is not necessary in our case, since we know that $x_0^c =1/2$, but the left panel of Fig.\ \ref{fig:3} shows that the threshold could  easily be determined if it were not known.  Of course, the validity of our scaling assumption (\ref{scal}) depends on whether we can determine the critical  exponent $\nu_\rho$ such that the curves for different values of $M$ `collapse' into a single curve, viz., the scaling function $h_N(u)$  \cite{Binder_1985,Privman_1990}. In fact, the right panel of Fig.\ \ref{fig:3} shows that the data collapse with $\nu_\rho=2$ is exceptionally good.

\begin{figure}[th] 
\centering
 \includegraphics[width=1\columnwidth]{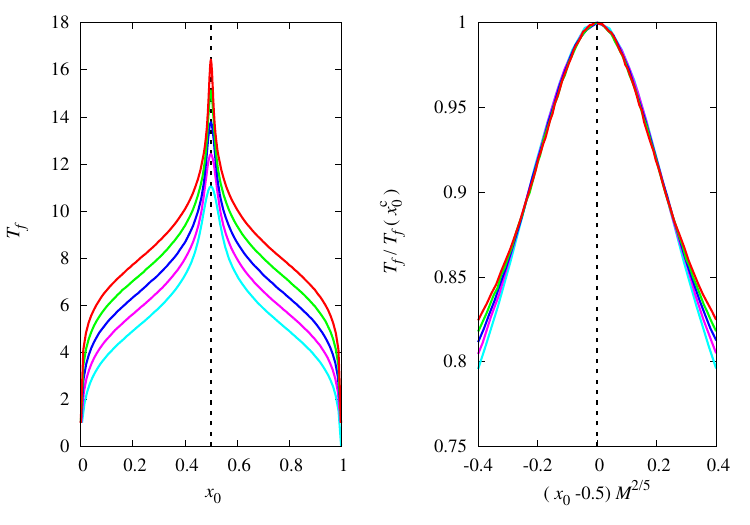}  
\caption{Mean time for fixation of one of the opinions  for play groups of size $N=4$ as a function of the initial proportion of individuals of type A (left panel). The  populations  sizes are  (from bottom to top) $M=200, 400, 800, 1600$, and $3200$.  
The right panel shows the ratio $T_f/T_f(1/2)$ as a function of the scaled variable $v = (x_0 - 1/2) M^{2/5}$. 
The vertical dashed lines indicate the threshold in the deterministic limit $M\to \infty$. The bias is  $\kappa=1/2$.
 }  
\label{fig:4}  
\end{figure}

\begin{figure}[th] 
\centering
 \includegraphics[width=1\columnwidth]{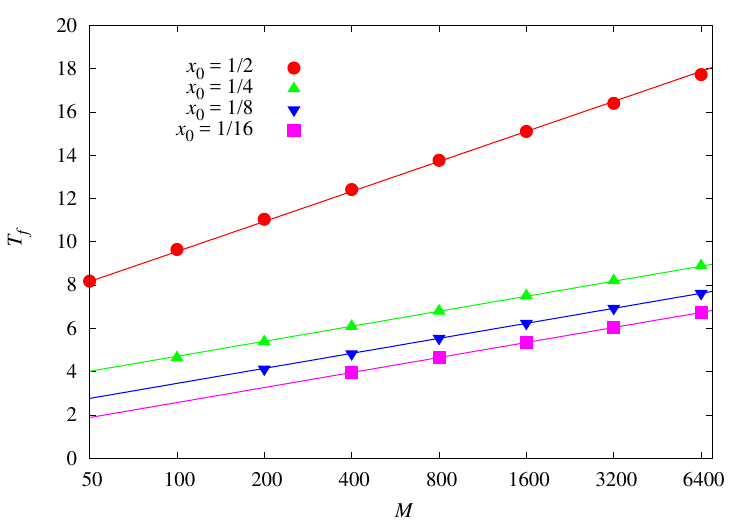}  
\caption{Mean time for fixation of one of the opinions  for play groups of size $N=4$ as a function of the population size $M$ for $x_0=1/2, 1/4, 1/8$ and $1/16$ as indicated.  The solid lines are the fits   $T_f =   \ln(1.40 M ^2) $ for $x_0 =1/2$, $T_f =   \ln (1.11 M)  $ for $x_0 =1/4$,   $T_f =   \ln(0.32 M) $ for $x_0 =1/8$ and $T_f =   \ln(0.13 M) $ for $x_0 =1/16$.  The bias is  $\kappa=1/2$.
 }  
\label{fig:5}  
\end{figure}

Figure \ref{fig:4} shows the mean time to fixation of either opinion A  or opinion B. There are two obvious features in this figure. First, as expected, $T_f$ is invariant to the transformation of $x_0$ into $1-x_0$, which simply amounts to swapping the labels of the opinions.  Second, $T_f$ increases very slowly  with increasing population size, suggesting a logarithmic dependence on  $M$. Indeed, Fig.\ \ref{fig:5} shows that this is the case, and that $T_f$ diverges faster at the threshold $x_0^c =1/2$ than at the other values of $x_0$. More specifically, $T_f \sim \ln M^2$ at the threshold, while $T_f \sim \ln M$ away from it.  This means that the width of the peak must decrease with increasing $M$. To quantify this effect, we assume the scaling relation
\begin{equation}\label{scalt}
T_f  = T_f (x_0^c)  g_N \left [ (x_0 - x_0^c)M^{1/\nu_T} \right ], 
\end{equation}
where $\nu_T > 0$ is a critical exponent and $g(v)$ with  $v=  (x_0 - x_0^c)M^{1/\nu_T}$  is a scaling function such that $g_N(0)=1$. As shown in the left panel of Fig. \ref{fig:4}, the relative mean fixation time in the vicinity of the threshold is well described by the exponent $\nu_T = 5/2$, which is surprisingly different from $\nu_\rho$.

%-----------------------------------------------------
%-----------------------------------------------------

%-----------------------------------------------------
%-----------------------------------------------------

Now we consider the effect of the group size $N$ on $\rho$ and $T_f$. As expected, we have verified that the critical exponents $\nu_\rho$ and $\nu_T$ do not depend on $N$ since these exponents reflect universal features of the opinion dynamics, such as  the procedure for selecting  individuals in a play group. For example,  if the individuals are fixed at the sites of a lattice and the play groups are formed by choosing their nearest neighbors, then the critical exponents  are expected to change.  Moreover,  for  $\kappa=1/2$ we have $x_0^c=1/2 $ regardless of the value of $N$. 

\begin{figure}[th] 
\centering
 \includegraphics[width=1\columnwidth]{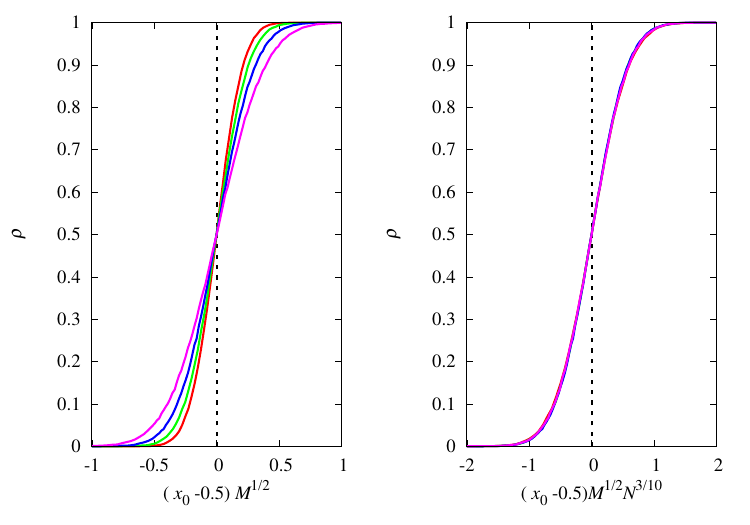}  
\caption{Probability of fixation of opinion A for populations  of size $M=3200$ as a function of the of the scaled variable $(x_0 - 1/2) M^{1/2}$ (left panel)  for play groups of sizes  (from left to right at $\rho=0.2$) $N=4, 8, 16$, and $32$. The right panel shows $\rho$ as a function of the double-scaled variable $(x_0 - 1/2) M^{1/2}N^{3/10}$. The vertical dashed lines indicate the threshold in the deterministic limit $M\to \infty$. The bias is  $\kappa=1/2$.
 }  
\label{fig:6}  
\end{figure}

  However, the scaling function $h_N(u)$ (and hence $\rho$)  depends on $N$, as shown in Fig.\ \ref{fig:6}. In fact, for fixed and large $M >N$, $h_N(u)$ becomes steeper as $N$ increases, signaling a threshold phenomenon in the limit $N \to \infty$.  More precisely, for $N$ fixed, we first take the limit $M \to \infty$ and $x_0 \to 1/2$ so that the scaled variable $u$ is finite, and then we take the limit $N \to \infty$.  Note that although the data for $h_N(u)$ were presented for $M=3200$, this scaling function is effectively independent of $M$, provided it is not too small, as shown in Fig.\ \ref{fig:3}. Our results (see right panel of Fig.\ \ref{fig:6}) show that close to the threshold $x_0^c = 1/2$ the probability of fixation of opinion A  is very accurately described  by the scaling  form
\begin{equation}\label{scal2}
\rho =  h \left [ (x_0 -1/2)M^{1/2} N^{3/10}  \right ], 
\end{equation}
where the scaling function $h$ does not depend on $M$ or $N$.  Note that the increase in the steepness of $\rho$ in the transition region is expected with increasing group size $N$: the smaller $N$, the greater the probability that the focal individual will copy a model individual that is a minority type in the population at large.

%-----------------------------------------------------
%-----------------------------------------------------

%-----------------------------------------------------
\begin{figure}[th] 
\centering
 \includegraphics[width=1\columnwidth]{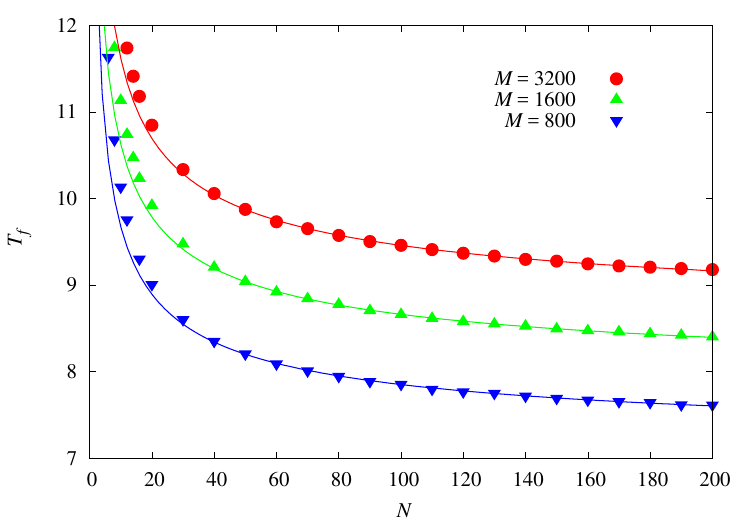}  
\caption{Mean time for fixation of one of the opinions  as a function of the  play group size for  populations of  size $M=800, 1600$ and $3200$ as indicated. The solid lines are the fits   $T_f =   8.43/ N^{1/2}  + 7.00$ for $M =800$,  $T_f =  9.09/ N^{1/2}  + 7.75 $ for $M=1600$ and $T_f =   9.99/ N^{1/2}  + 8.46$ for $M=3200$.  The initial proportion of  type A individuals is $x_0 = 1/2$ and the  bias is  $\kappa=1/2$.
 }  
\label{fig:7}  
\end{figure}
%-----------------------------------------------------

As shown in Fig.\ \ref{fig:7}, another effect of increasing group size is a decrease in the time to fixation of one of the opinions. In fact, $T_f$ decreases as $1/N^{1/2}$ to a value that depends on the size of the population. We recall that in this paper we consider the scenario $M \gg N$, which recovers the framework of the replicator equation in the infinite population limit.

%
%-----------------------------------------------------
\section{The effect of inflexible individuals}\label{sec:Inf} 
%-----------------------------------------------------
%
A good way to illustrate the differences between the $N$-player game model and Galam's model of opinion dynamics is to consider the effect of inflexible individuals, i.e.,  individuals who never change their opinions \cite{Galam_2007,Martins_2008}.  Here we briefly discuss this problem for the case where the population is infinite   and the inflexible individuals hold only opinion A.  Let $x$ and $y$ represent the proportions of individuals holding opinions A and B, respectively, but now consider that a fraction $z$ of the infinite population consists of inflexible individuals, so that $x+y+z=1$.  Individuals who can change their opinions are called floaters \cite{Galam_2012}. The payoff of an inflexible individual is $w$ regardless of the composition of the play group. This guarantees  that an inflexible individual will never switch to opinion B. The average payoffs can be written more compactly  in terms of the proportion $y$ of floaters with opinion B,
\begin{eqnarray}\label{piiA}
\pi^A  (y) &  =  &  w - 
(w-l)  \sum_{i=0}^{N/2-2} \binom{N-1}{i} (1-y)^{i} y^{N-1-i}  \nonumber \\
&  & -(1-\kappa) (w-l) \binom{N-1}{N/2-1} (1-y)^{N/2-1} y^{N/2} ,  \nonumber \\
\end{eqnarray}
\begin{eqnarray}\label{piiB}
\pi^B  (y)  &  =  &  w - 
(w-l)  \sum_{j=0}^{N/2-2} \binom{N-1}{j} (1-y)^{N-1-j} y^{j} \nonumber \\
&  & - \kappa  (w-l) \binom{N-1}{N/2} (1-y)^{N/2} y^{N/2-1} ,
\end{eqnarray}
and 
\begin{equation}\label{piiI}
\pi^I  (y) = w .
\end{equation}

%-----------------------------------------------------
\begin{figure}[t] 
\centering
 \includegraphics[width=1\columnwidth]{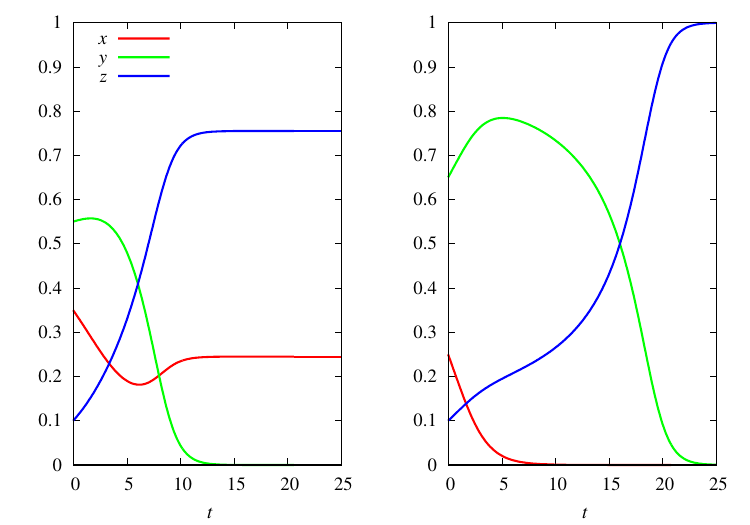}  
\caption{Numerical solution of the replicator equations (\ref{repx})-(\ref{repz})  for play group size $N=4$ and initial conditions $x_0=0.35$, $y_0=0.55$, $z_0=0.1$ (left panel) and  $x_0=0.25$, $y_0=0.65$, $z_0=0.1$ (right panel). The bias is  $\kappa=1/2$.
 }  
\label{fig:8}  
\end{figure}
%----------------------------------------------------- 
%
The proportions of the different types of individuals obey the replicator equations
\begin{eqnarray}
\frac{dx}{dt}   & = &  x   \left [ \pi^A (y) - \bar{\pi}  \right ]  \label{repx} \\
\frac{dy}{dt}   & = &  y   \left [ \pi^B (y) - \bar{\pi}  \right ]  \label{repy} \\
\frac{dz}{dt}   & = &  z   \left [ \pi^I (y) - \bar{\pi}  \right ] \label{repz}
\end{eqnarray}
where $\bar{\pi}  = x \pi^A (y) + y \pi^B (y) + z \pi^I (y)$ is the average payoff of the population, which ensures that $d(x+y+z)/dt = 0$.  As before, the factor $w-l$ can be absorbed in a rescaling of the variable $t$, so the solution of these equations depends only on the size of the play group $N$, the bias $\kappa$, and the initial conditions $x_0 = x(0)$, $y_0 = y(0)$ and $z_0 = z(0)$, with $x_0+y_0+z_0=1$. Only the case $z_0 > 0$ needs to be considered here. We find that $y=0$ in the equilibrium regime, i.e., the presence of even a small fraction of inflexible individuals with opinion A causes opinion B to disappear. Whether or not floaters with opinion A are present in the equilibrium regime depends on the initial conditions, as shown in Fig.\ \ref{fig:8}. In fact, if opinion B is the majority in the initial steps of the dynamics (i.e., $y_0 > 0.5$), then the floaters with opinion A may eventually disappear if the proportion of inflexible individuals does not grow fast enough to reach a majority ($z>0.5$) in a short time. Note that in Galam's model the fraction of inflexible individuals does not change over time, while in the $N$-player game this fraction always increases over time due to their superior payoff, provided there are floaters with opinion B in the population. When these floaters disappear, the dynamics freezes at the current proportions of inflexible individuals and floaters with opinion A, as shown in the left panel of Fig.\ \ref{fig:8}.  In a more realistic scenario where there are also inflexible individuals holding opinion B, our model predicts a polarized equilibrium regime characterized by the presence of only inflexible individuals.

We note that individuals who never change their opinions -- extremists or inflexible individuals \cite{Galam_2007} -- are also referred to as  
zealots in the opinion dynamics literature.  In one and two dimensional lattices with nearest neighbor interactions (voter model), a single zealot can influence an infinite group of voters to adopt her opinion, but in higher dimensions the zealot cannot influence all individuals \cite{Mobilia_2003}. These results have been applied to the study of the diffusion of innovations in lattice models, where the innovator is considered a zealot \cite{Tilles_2015}.   Of course, since in our finite population version of the majority-vote model a floater will only change her opinion if she is in the minority in the play group, it takes a minimum of $N/2$ zealots to achieve this change and sway the entire population to the zealots' opinion. In addition, by promoting coexistence, zealotry has a surprising effect on the dynamics of evolutionary games used to study biodiversity \cite{Verma_2015, Szolnoki_2016}.

%
%-----------------------------------------------------
\section{Conclusion}\label{sec:conc} 
%-----------------------------------------------------
%

Opinion dynamics, although a subject originating and mostly belonging to social psychology \cite{Latane_1981}, has been a topic of great interest and activity in the statistical physics community \cite{Castellano_2009,Sznajd_2019}. Similarly, evolutionary game theory, which was introduced to understand animal behavior in conflict scenarios \cite{Maynard_1973,Maynard_1982}, has become a major research topic in statistical physics \cite{Perc_2017,Wang_2023}. Here we combine these two approaches to develop an $N$-player game model of (binary) opinion formation that provides an alternative, but not equivalent, formulation of Galam's majority-vote model \cite{Galam_2012}. Evolutionary game theory models of opinion formation have been considered before (see, e.g., \cite{Ding_2010,Gargiulo_2012,Yang_2016,Li_2022}), but have been limited to pairwise interactions between players (i.e., $2$-player games). 
Our $N$-player game formulation allows the full use of the replicator equation formalism to study the infinite population limit of the opinion dynamics model. In particular, we find that one of the two competing opinions, say opinion A, becomes fixed in the population if its initial frequency $x_0$ is greater than a threshold $x^*$, which depends on the population bias $\kappa$ toward opinion A. In the unbiased case $\kappa=1/2$, we find $x^*=1/2$, as expected.

The connection between the replicator equation and the stochastic imitation dynamics for finite populations \cite{Traulsen_2005} prompts a finite-size scaling analysis of the finite population opinion  model  with some surprising results. We find that the sharpness of the threshold  for the probability of fixation of opinion A  increases with $M^{1/2}$, where $M$ is the population size, and with $N^{3/10}$, where $N$ is the play group size (see Fig.\ \ref{fig:6}), provided that $M \gg N$.  In addition, we find that the mean time to fix one of the opinions increases with $\ln M$ away from the threshold and with $\ln M^2$ at the threshold. Furthermore, the width of the peak of the mean fixation time at $x^*$ decreases with $M^{-2/5}$ (see Fig. \ref{fig:4}) and is not affected by the play group size $N$.  Although these critical exponents were obtained for the unbiased case, we expect them to hold for other values of $\kappa$ as well. The remarkable quantitative understanding of the stochastic imitation dynamics offered by finite-size scaling, summarized in Eqs. (\ref{scalt}) and (\ref{scal2}),  attests to the usefulness of statistical physics in addressing such interdisciplinary problems.

In our $N$-player game, individuals who hold the majority opinion in their play groups have a larger payoff than individuals who hold the minority opinion. Thus, the payoffs are not associated with the opinions themselves.  Individuals have a tendency to copy the opinions of others who have greater payoffs, leading to a more nuanced scenario than the direct application of the majority rule to effect the simultaneous and certain switching of opinion by the minority side in a discussion group, as is done in Galam's model. For example, in our scenario, inflexible individuals (extremists) increase their frequency in the population as long as there are floaters with the opposite opinion, leading to complete polarization when there are extremists on both sides. It would be interesting to explore the variety of strategies people use to influence others' opinions within the $N$-player game theoretical framework.

\bigskip

\acknowledgments
JFF is indebted to Mauro Santos for the many discussions on evolutionary game theory.
JFF is partially supported by  Conselho Nacional de Desenvolvimento Ci\-en\-t\'{\i}\-fi\-co e Tecnol\'ogico -- Brasil (CNPq) -- grant number 305620/2021-5.  JMS is supported by CNPq grant number 126343/2023-3.

%\section*{References}


\begin{thebibliography}{}

\bibitem{Asch_1955}
 S.E. Asch, 
 Opinions and social pressure,
  Sci. Am. 193 (5) (1955) 31--35,
  https://doi.org/10.1038/scientificamerican1155-31.
  
  \bibitem{Liggett_1985}
T. M. Liggett, 
Interacting Particle Systems,
Springer, New York, 1985.

\bibitem{Castellano_2009}
C. Castellano, S. Fortunato, V. Loreto, 
Statistical physics of social dynamics, 
Rev. Mod. Phys. 81 (2) (2009) 591--646,
https://doi.org/10.1103/RevModPhys.81.591.

\bibitem{Galam_2012}
S. Galam,
Sociophysics:  A Physicist's Modeling of Psycho-political Phenomena, 
Springer, New York, 2012.

\bibitem{Sznajd_2019}
A. J\k{e}drzejewski, K. Sznajd-Weron,
Statistical Physics Of Opinion Formation: Is it a SPOOF?,
C. R. Phys. 20 (4) (2019) 244--261,
https://doi.org/10.1016/j.crhy.2019.05.002

\bibitem{Boyd_1985}
 R. Boyd, P.J. Richerson, 
 Culture and the evolutionary process,
 University of Chicago Press, Chicago, 1985.

 \bibitem{Perc_2017}
 M. Perc, J.J. Jordan, D.G. Rand, Z. Wang, S. Boccaletti, A. Szolnoki, 
Statistical physics of human cooperation,
Phys. Rep. 687 (2017) 1--51,
https://doi.org/10.1016/j.physrep.2017.05.004.

 \bibitem{Wang_2023}
C. Xia, J. Wang,  M. Perc,  Z. Wang, 
Reputation and reciprocity,
Phys.  Life Rev. 46 (2023) 8--45,
https://doi.org/10.1016/j.plrev.2023.05.002.

\bibitem{Hofbauer_1998}
J. Hofbauer, K.  Sigmund,  
Evolutionary Games and Population Dynamics,
Cambridge University Press, Cambridge, 1998.

\bibitem{Nowak_2006}
M.A. Nowak,   
Evolutionary Dynamics: Exploring the Equations of Life, 
Belknap Press,  New York, 2006.

 \bibitem{Traulsen_2005}
A. Traulsen,  J.C. Claussen, C.  Hauert, 
Coevolutionary Dynamics: From Finite to Infinite Populations,
Phys. Rev. Lett.  95 (23) (2005) 238701,
 https://doi.org/10.1103/PhysRevLett.95.238701.
 
 \bibitem{Sandholm_2010}
W.H. Sandholm,
Population Games and Evolutionary Dynamics, 
MIT Press, Cambridge,  2010.



%%%%%%%%%%%%%%%%%%%%%%%%%%

%--- Galam's model

\bibitem{Galam_2000}
S. Galam, 
Real space renormalization group and totalitarian paradox of majority rule voting, Physica A 285 (1-2) (2000) 66--76, 
https://doi.org/10.1016/S0378-4371(00)00272-7.

\bibitem{Galam_2002}
S. Galam,
Minority opinion spreading in random geometry,
Eur. Phys. J. B 25  (2002) 403--406,
 https://doi.org/10.1140/epjb/e20020045.
 
 \bibitem{Galam_2005a}
S. Galam,
Local dynamics vs. social mechanisms: A unifying frame,
Europhys. Lett. 70 (6) (2005)  705,
https://doi:10.1209/epl/i2004-10526-5.



\bibitem{Galam_2005b}
S. Galam, 
Heterogeneous beliefs, segregation, and extremism in the making of public opinions, 
Phys. Rev. E 71 (4) (2005) 046123, 
https://doi.org/10.1103/PhysRevE.71.046123.



\bibitem{Oliveira_1992}
M.J. de Oliveira,  
Isotropic majority-vote model on a square lattice,
J Stat Phys 66 (1992) 273--281,
 https://doi.org/10.1007/BF01060069.

\bibitem{Peres_2010}
 L.R. Peres,  J.F. Fontanari,
Statistics of opinion domains of the majority-vote model on a square lattice,
Phys. Rev. E  82 (4)  (2010) 046103,
https://doi.org/10.1103/PhysRevE.82.046103.

 \bibitem{Tome_2023}
 T. Tom\'e, C.E. Fiore, M.J. de Oliveira,
 Stochastic thermodynamics of opinion dynamics,
 Phys. Rev. E 107 (6) (2023) 064135,
 https://doi.org/10.1103/PhysRevE.107.064135.


\bibitem{Cooper_1998}
R. Cooper,
 Coordination Games, 
 Cambridge University Press, Cambridge, 1998.
 
 \bibitem{Zheng_2007}
D.F. Zheng,  H.P. Yin, C.H. Chan, P.M.  Hui,
Cooperative behavior in a model of evolutionary snowdrift games with $N$-person interactions, 
Europhys. Lett.  80 (1)  (2007) 18002, 
 https://doi.org/10.1209/0295-5075/80/18002.
 
  \bibitem{Fontanari_2024}
 J.F. Fontanari,
 Imitation dynamics and the replicator equation,
Europhys. Lett.  146  (2024) 47001,
 https://doi.org/10.1209/0295-5075/ad473e.
 
 
 
  \bibitem{Hardin_1968}
J. Hardin,
The tragedy of the commons, 
Science 162  (1968) 1243--1248,
 https://doi.org/10.1126/science.162.3859.1243.
 
 \bibitem{Ostrom_1990}
E. Ostrom, 
Governing the Commons: The Evolution of Institutions for Collective Active,
Cambridge University Press, Cambridge, 1990.
 


\bibitem{Blackmore_2000} 
S. Blackmore, 
The Meme Machine,
Oxford University Press,  Oxford, 2000.

\bibitem{Fontanari_2014}
J.F. Fontanari, 
 Imitative Learning as a Connector of Collective Brains,
PLoS ONE 9  (2014)  e110517,
https://doi.org/10.1371/journal.pone.0110517.


  
   \bibitem{Twain_2014}
 M. Twain,
 Following the Equator,
 Wanderlust, Denver,  2014.



\bibitem{Strogatz_2014}
S.H. Strogatz,  
 Nonlinear Dynamics and Chaos: With Applications to Physics, Biology, Chemistry, and Engineering,
 Westview Press, New York,  2014.

\bibitem{Serva_2005}
M. Serva,
 On the genealogy of populations: trees, branches and offspring,
J. Stat. Mech. 2005  (2005) P07011,
 https://doi.org/10.1088/1742-5468/2005/07/P07011.

 \bibitem{Binder_1985}
 K. Binder, 
 The Monte Carlo method for the study of phase transitions: A review of some recent progress,
 J. Comp. Phys. 59 (1985) 1--55,
 https://doi.org/10.1016/0021-9991(85)90106-8.

\bibitem{Privman_1990}
 V. Privman, 
Finite-Size Scaling and Numerical Simulations of Statistical Systems,
 World Scientific, Singapore, 1990.
 
\bibitem{Galam_2007}
 S. Galam, F. Jacobs,
The role of inflexible minorities in the breaking of democratic opinion dynamics,
Physica A 381 (2007) 366--376,
https://doi:10.1016/j.physa.2007.03.034.

\bibitem{Martins_2008}
A.C.R. Martins, 
Mobility and Social Network Effects on Extremist Opinion, 
Phys. Rev. E 78 (3) (2008) 036104, 
https://doi.org/10.1103/PhysRevE.78.036104.


\bibitem{Mobilia_2003}
M. Mobilia,
Does a Single Zealot Affect an Infinite Group of Voters?
Phys. Rev. Lett. 91 (2) (2003) 028701, 
https://doi.org/10.1103/PhysRevLett.91.028701.


\bibitem{Tilles_2015}
P.F.C. Tilles,  J.F. Fontanari,
Diffusion of innovations in Axelrod's model,
 J. Stat. Mech. (2015) P11026,
https://doi.org/10.1088/1742-5468/2015/11/P11026.


\bibitem{Verma_2015}
G. Verma, K. Chan, A. Swami, 
Zealotry promotes coexistence in the rock-paper-scissors model of cyclic dominance,
Phys. Rev. E 92 (5) (2015) 052807,
https://doi.org/10.1103/PhysRevE.92.052807.

\bibitem{Szolnoki_2016}
A. Szolnoki, M. Perc.
Zealots tame oscillations in the spatial rock-paper-scissors game,
Phys. Rev. E 93 (6) (2016) 062307, 
https://doi.org/10.1103/PhysRevE.93.062307.


%--- Discussion

\bibitem{Latane_1981}
B. Latan\'e,  
The psychology of social impact,
Am. Psychol. 36 (4) (1981) 343--356, 
https://doi.org/10.1037/0003-066X.36.4.343



\bibitem{Maynard_1973}
J. Maynard Smith,  G.R. Price, 
 The logic of animal conflict, 
 Nature 246 (1973) 15--18,
 https://doi.org/10.1038/246015a0.
 
\bibitem{Maynard_1982}
J. Maynard Smith, 
Evolution and the Theory of Games, 
Cambridge University Press, Cambridge, 1982.




%%%%%


\bibitem{Ding_2010}
F.  Ding, Y. Liu, B. Shen, X.-M.  Si,
An evolutionary game theory model of binary opinion formation,
Physica A 389 (8) (2010)  1745--1752,
https://doi.org/10.1016/j.physa.2009.12.028.


\bibitem{Gargiulo_2012}
F. Gargiulo, J.J.  Ramasco,
Influence of Opinion Dynamics on the Evolution of Games,
PLoS ONE 7 (11) (2012) e48916,
https://doi.org/10.1371/journal.pone.0048916.

\bibitem{Yang_2016}
H.-X. Yang,
A consensus opinion model based on the evolutionary game,
Europhys. Lett.   115 (4) (2016) 40007,
https://doi.org/10.1209/0295-5075/115/40007.

\bibitem{Li_2022}
Z. Li, X. Chen, H.-X. Yang, A. Szolnoki,
Game-theoretical approach for opinion dynamics on social networks,
Chaos 32 (2022) 073117,
https://doi.org/10.1063/5.0084178. 



 
\end{thebibliography}
\end{document}